\documentclass[12pt]{article}
\usepackage[dvips]{color}
\usepackage{amsmath}
\usepackage{graphicx}
\usepackage{subfigure}
\usepackage{epsfig}
\usepackage{amsmath}
\usepackage{cite}
\usepackage{color}
\usepackage{subfigure}

\begin{document}
\begin{center}
\Large{\bf Cosmic evolution of the logarithmic $f(R)$ model and the dS swampland conjecture}\\
\small \vspace{1cm} {\bf J. Sadeghi$^{a,d}$\footnote {Email:~~~pouriya@ipm.ir}},
 {\bf B. Pourhassan$^{b,c,d}$\footnote {Email:~~~b.pourhassan@du.ac.ir}},
{\bf S. Noori Gashti$^{a}$\footnote {Email:~~~saeed.noorigashti@stu.umz.ac.ir}},
{\bf E. Naghd Mezerji$^{a}$\footnote {Email:~~~e.n.mezerji@stu.umz.ac.ir}},
{\bf A. Pasqua$^{e}$\footnote {Email:~~~toto.pasqua@gmail.com}}
\\
\vspace{0.5cm}$^{a}${Department of Physics, Faculty of Basic Sciences,\\
University of Mazandaran P. O. Box 47416-95447, Babolsar, Iran}\\
\vspace{0.5cm}$^{b}${School of Physics, Damghan University, P. O. Box 3671641167, Damghan, Iran.}\\
\vspace{0.5cm}$^{c}${Physics Department, Istanbul Technical University, Istanbul 34469, Turkey.}\\
\vspace{0.5cm}$^{d}${Canadian Quantum Research Center 204-3002 32 Ave Vernon, BC V1T 2L7 Canada.}\\
\vspace{0.5cm}$^{e}${Department of Physics, University of Trieste, Via Valerio, 2 34127 Trieste, Italy.}\\
\small \vspace{1cm}
\end{center}
\begin{abstract}
In this paper, we study the inflationary scenario in logarithmic $f(R)$ gravity, where the rate of inflation roll is constant. On the other hand,  our gravitational $f(R)$ model is a polynomial plus a logarithmic term. We take advantage of constant-roll conditions and investigate the cosmic evolution of the logarithmic $f(R)$ gravity. Therefore, we plot some figures such as the scalar spectrum index $n_{s}$ and tensor-to-scaler ratio $r$ concerning $n$, $\beta$ and model's constant parameters, i.e., $\alpha$, $\theta$ and $\gamma$ respectively. Also, we obtain the potential by using the constant roll condition. We know that the potential value obtained with this condition has an exact value. Next, we challenge it with refined swampland conjecture with respect to the Planck data. Finally, we compare our results with the experimental data, especially Planck 2018.
\end{abstract}
\newpage

\section{Introduction}
Inflation models have been proposed to address and solve several problems such as the horizon, flatness, and the absence of magnetic monopoles. This theory states that our early universe has gone through a period of accelerated expansion. A scalar field could be responsible for this positive acceleration; in fact, their quantum oscillations were the seeds of the formation of structures. Due to their inflation potential and their interactions, different inflation models have been used in the literature. The inflationary paradigm in cosmology is one of the plausible scenarios that describe the early evolution of the universe; indeed, the description of the inflationary period in terms of the slow-rolling scalar field and there are many studies in the literature that explicitly describe single scalar inflation.\\
Most researchers agree that one of the most important cosmology problems is the accurate description of the early universe. Our universe has gone through a period of inflation that has led to its expansion. Inflation of universe can be described by several theory as usual gravitational or modified $f(R)$ gravitational models \cite{1,2}. Modified gravity appears in different forms and generally plays an important role in describing the universe's evolution \cite{3,4,5,6}. In particular, a large number of phenomena related to different stages of evolution related to the present university can be investigated by using modified gravity theories\cite{7}. Among the various theories related to modified gravity, $f(R)$ gravity is one of the best and most common theories related to gravity and its simplicity of presentation and its high concept about the universe and its properties. As we know, the theoretical framework of $f(R), $ gravity lead us to find the unifying description of acceleration periods associated with the universe,  as an early-time and late-time acceleration era\cite{8,9,10}. We note here many approaches have been proposed in the literature for the investigation of inflation. In that case, we have different methods such as scalar field potentials, modified gravity, etc. Such inflation describes the early evolution of the universe. So, due to inflationary potentials, different models of inflation have been used in the literature. Generally, one can say that the slow-roll conditions examine the inflationary era. The inflationary scenario have been studied from different structure of modified gravity such as $f(R)$, $f(R, T)$, also with various methods such as constant-roll, slow-roll, ultra-slow-roll conditions\cite{11,12,13,14,15,16,166,17,18,19,20,21,22,23,24,25,27,28,29,30,31,32,33,34,35,36,37,38,39,jhap,jhap2} Recently, a lot of work has been done in the term of weak gravity conjecture (WGC) in relation to the swampland, landscape, and trans-Planckian censorship conjecture (TCC)\cite{26,266,2666,311,3111,110,111,112,113,114,115,116,117,118,119,120,121,122,123}. In this paper, we are going to investigate the constant-roll evolution of modified $f(R)$ gravity, which is polynomial plus a logarithmic term. Here,  we want to analyze and evaluate the scalar index spectrum $n_{s}$  and tensor-to-scalar ratio with respect to  $n$ and $\beta$ of this $f(R)$ gravity. We briefly explain and then examine our inflationary model. Therefore, in section 2, we will first introduce the concepts and relations related to the gravitational model's evolution. In section 3, we introduce the modified gravitational model, and we examine some corresponding relations discussed in the previous Section. In this section, we also analyze the above model with some different figures. In section 4, we investigate the potential of our logarithmic inflation model by applying the constant roll-condition, and we challenge it with refined swampland conjecture. Finally in the last section, we will explain the paper results.

\section{$f(R)$ gravity and constant-roll evolution}
In this Section, we assume that a constant-roll era has occurred during the period of inflation. The inflationary paradigm of constant-roll has been used in the content of scalar-tensor theories\cite{32,33,34,35,36,37,38,39} as well as in the generalized content of $f(R)$ modified gravity\cite{40,41,42} and many have examined it in previous works. We first give a brief explanation about $f(R)$ gravity, and then we study our inflationary model\cite{b,50}. We consider the action which is given by,

 \begin{equation}\label{1}
S=\int d^{4}X \sqrt{-g_{j}}\frac{f(R_{j})}{2},
 \end{equation}

The gravitational field equations for the background metric of Friedman-Lemitre-Robertson-Walker are presented as.

 \begin{equation}\label{2}
ds^{2}=-dt^{2}_{j}+\alpha^{2}_{j}(t)(dx^{2}+dy^{2}+dz^{2}),
 \end{equation}
The above relations were in the Jordan frame, but we can transfer these relations to the Einstein framework by a conformal transformation such as $g^{E}_{\mu\nu}=Fg^{j}_{\mu\nu}$. So using these conformal transformations, the above action which is given by,

 \begin{equation}\label{3}
S=\int d^{4}X \sqrt{-g_{E}}(f(R)_{E}-V(\phi)).
 \end{equation}

Which, like the above relations $j$ was the symbol of the Jordan frame, here subscript $E$ denotes the Einstein frame. also one can obtained,

 \begin{equation}\label{4}
V(\phi)= \frac{RF-f}{2F^{2}},
 \end{equation}

\begin{equation}\label{5}
F=\frac{df}{dR}.
 \end{equation}

We want to describe the modified $f(R)$ gravity, which has an important role in describing dark energy and cosmic acceleration and introducing our gravitational model.

The most natural extension constant-roll condition considered in most works is usually in the following form:

 \begin{equation}\label{6}
-\frac{\ddot{H}}{2H\dot{H}}= \beta,
 \end{equation}

where $\beta$ is a constant parameter that can have positive or negative values. Eq. (\ref{6}) can be written by,

 \begin{equation}\label{7}
-\frac{\ddot{\phi}}{H \dot{\phi}}=\beta,
 \end{equation}

Moreover, the second slow-roll condition  is $\eta\sim-\frac{\ddot{H}}{2H\dot{H}}$. We assume that we have a theory described by $f(R)$ gravity and that the background is a flat FRW metric. According to variation $f(R)$ gravity concerning metric, one can have the following equation of motion,

 \begin{equation}\label{8}
3F_{R}H^{2}=\frac{F_{R}R-F}{2}-3H\dot{F_{R}},
 \end{equation}
 \begin{equation}\label{9}
-2F_{R}\dot{H}=\ddot{F}-H\dot{F},
 \end{equation}

where $F_{R}=\frac{\partial F}{\partial R}$ and a dot denotes a derivation with respect to $t$. The dynamics of $f(R)$ gravity inflation with the four inflation indicators $\epsilon_{i}, i=1...4$ expressed as follows\cite{43,44,45,46,47,48},

\begin{equation}\label{10}
\epsilon_{1}=-\frac{\dot{H}}{H^{2}},\hspace{10pt} \epsilon_{2}=0,\hspace{10pt}\epsilon_{3}=\frac{\dot{F_{R}}}{2HF_{R}}, \hspace{10pt}\epsilon_{4}=\frac{\dot{E}}{2HE},
 \end{equation}

where $E=\frac{3\dot{F_{R}^{2}}}{2\kappa^{2}}$. To calculate the tensor to scalar ratio $r,$ it is necessary to calculate the $Q_{s}$, which is also expressed as follows.

\begin{equation}\label{11}
Q_{s}=\frac{E}{F_{R}H^{2}(1+\epsilon_{3})^{2}},
 \end{equation}

The spectral index of curvature perturbations $n_{s}$, $\dot{\epsilon_{i}}\simeq0$, one can obtain,\cite{44,45,46},

\begin{equation}\label{12}
n_{s}=4-2\sqrt{\frac{1}{4}+\frac{(1+\epsilon_{1}-\epsilon_{3}+\epsilon_{4})(2-\epsilon_{3}+\epsilon_{4})}{(1-\epsilon_{1})^{2}}}
 \end{equation}

The above relation is a general one that is true anyway. According to tensor to the scalar ratio in the content of modified $f{R} $ gravity theory, we have

\begin{equation}\label{13}
r=\frac{8\kappa^{2}Q_{s}}{F_{R}},
 \end{equation}

concerning Eq. (\ref{11}), and for the case of a $f(R)$ gravity, the tensor-to-scalar ratio one can obtain,

\begin{equation}\label{14}
r=\frac{48\epsilon_{3}^{2}}{(1+\epsilon_{3})^{2}},
 \end{equation}

Now with respect to Eq. (\ref{6}), which affect Eq. (\ref{10}), which can be written as follows \cite{41},

\begin{equation}\label{15}
\epsilon_{1}=-\frac{\dot{H}}{H^{2}}, \hspace{10pt} \epsilon_{2}=0, \hspace{10pt} \epsilon_{3}=\frac{\dot{F_{RR}}}{2HF_{R}}(24H\dot{H}+\ddot{H}), \hspace{10pt} \epsilon_{4}=\frac{F_{RRR}}{HF_{R}}\dot{R}+\frac{\ddot{R}}{H\dot{R}},
 \end{equation}

where $F_{RR}$ and $F_{RRR}$ are $\frac{\partial^{2}F}{\partial R^{2}}$ and $\frac{\partial^{3}F}{\partial R^{3}} $ respectively. It can be seen that inflationary dynamics are related to $F (R)$ gravity. We are studying the modified $f(R)$ gravity; also, as we will show in the next Section, this model has successfully described the late acceleration period.

\section{Logarithmic $f(R)$ gravity}
The conditions and restrictions applied to gravitational models always lead to changes in the features of these models. For example, the constant-roll conditions change the durability of the $f(R)$ gravitational model \cite{41}. This Section wants to take a closer insight of our inflationary model as a modified $f(R)$ gravitational model with polynomial plus logarithmic terms and see what happens for the corresponding model. As you know, this kind of  $f(R)$ inflationary model has the following  form,

\begin{equation}\label{16}
f(R) = R + \alpha R^{2} + \theta R^{n} + \gamma R^{2}\ln\gamma R,
 \end{equation}

Here we note that the logarithmic $f(R)$ gravitational model describes neutron stars, the polynomial section describes cosmological models, and the gluon effects \cite{28,b,30}, where $n$, $\alpha$, $\theta$, and $\gamma$ are constant parameters, these coefficients are responsible for the dimensional problem of the corresponding model. It would be reasonable to assume that inflation was predominant in the early universe, and thus we could ignore the contributions of matter and
its radiation. According to the above relation, we have

\begin{equation}\label{01}
f'(R)=1+(2\alpha+\gamma)R+2\gamma R\ln\gamma R+n\theta R^{n-1}, \hspace{20pt} f''(R)=2\alpha+\gamma+2\gamma\ln \gamma R+n(n-1)\theta R^{n-2}.
\end{equation}

We can explain the importance of this model according to the above two equations. If the mentioned model the parameters $\alpha\neq\theta\neq0$ and only the parameter $\gamma$ is zero, the model that is examined in Ref. \cite{d}. With $\theta=0$, the model is reduced to the famous Starobinsky model. A special investigation with $\gamma=0$, and $n=4$ in Ref. \cite{e} has also been performed. The $f(R)$ also satisfies the conditions $f(0)=0$ that lead to a flat space-time without a cosmological constant. Also, the stability of this model is thoroughly investigated in\cite{f}. Another critical point is the review of this model by the article's authors in connection with Investigating the logarithmic form of the $f(R)$ gravity model from brane's perspective and swampland criteria, which has exciting results that you can see in more detail in Ref. \cite{g}. This model has also been studied in examining a specific type of traversable wormholes concerning various shape and redshift functions. Its exciting results have also been investigated in Ref. \cite{h}. In addition, its inflation model has been challenged with special conditions, i.e., slow-roll and weak gravity conjecture; the details of this study can be seen in Ref. \cite{26}. According to the above model, quantum stability conditions are obeyed by $f(R)$, and also according to the above equations, classical stability conditions also lead to,

\begin{equation*}\label{02}
f'(R)=1+(2\alpha+\gamma+2\gamma\ln \gamma R)R+n\theta R^{n-1}>0.
\end{equation*}

Now, we consider  Eq. (\ref{8}), which for the above inflationary model, it can be approximated as follows,

\begin{equation}\label{17}
 \begin{split}
&-3H^{2}\bigg[1+18(2+\beta)\alpha H^{2}+\theta6^{n-1}\big((2+\beta)H^{2}\big)^{n-1}n+12(2+\beta)H^{2}\gamma\ln\big(6(2+\beta)H^{2}\big)\gamma\bigg]\\ &+\frac{1}{2}\bigg[-6(2+\beta)\alpha H^{2}-36(2+\beta)^{2}H^{4}-\theta 6^{n}\big((2+\beta)H^{2}\big)^{n}-36(2+\beta)^{2}H^{4}\gamma\ln\big(6(2+\beta)H^{2}\big)\gamma\bigg]\\
&+6(2+\beta)\alpha H^{2}\bigg[1+18(2+\beta)H^{2}\theta6^{n-1}\big((2+\beta)H^{2}\big)^{n-1}n+12(2+\beta)H^{2}\gamma\ln\big(6(2+\beta)H^{2}\big)\gamma\bigg]\\
&-3H\bigg[1+36(2+\beta)\alpha H \dot{H}+12^{n-1}n \theta\big((2+\beta)H\dot{H}\big)^{n-1}+24(2+\beta)H \dot{H}\gamma\ln\big(12(2+\beta)H\dot{H}\big)\gamma\bigg]=0
 \end{split}
 \end{equation}

In the above calculations, we have used the constant-roll conditions, Eq. (\ref{6}). By performing a series of manipulations and straightforward calculations, and with a series of simplifications, we obtain the final relation by solving the differential equation for the Hubble parameter $H$. The general form is as follows,
\begin{equation}\label{18}
H(t)=-\frac{A}{B+C},
\end{equation}
where
\begin{eqnarray}\label{18-1}
C&=&\exp{\left(\frac{\gamma n\big[24+12\alpha\beta+\theta12^{n}(2+\beta)^{n}n\big]\big[t+144(2+\beta)^{2}\big(2^{2n+1}
\theta3^{n}(2+\beta)^{n}+3n\big)c_{1}\big]}{144(2+\alpha\beta)^{2}\gamma\big(2^{2n+1}3^{n}(2+\beta)^{n}\alpha+3n\big)}\right)},\nonumber\\
A&=&n\big[24+12\alpha\beta+\theta12^{n}(2+\beta)^{n}n\big]\gamma,\nonumber\\
B&=&24n+2\bigg[6\alpha\beta n+\theta6^{n}(2+\beta)^{n}\gamma\big(-36\alpha\beta(2+\beta)^{2}+(2+\beta)n-(1+\beta)n^{2}\big)\bigg].
\end{eqnarray}
Here,  we note that  $c_{1}$ is an arbitrary integration constant that is not affected in inflation dynamics. Using the above equation, i.e., the Hubble rate, you can easily get the slow-roll indices, $ \epsilon_{i}, i=1...4$. Then, using Eq. (\ref{15}), the indices of slow-roll for our inflation model will be following,

\begin{equation}\label{19}
\epsilon_{1}=\frac{\dot{H}}{H^{2}},
 \end{equation}
 and
\begin{equation}\label{20}
\epsilon_{2}=0,
 \end{equation}
 while
 \begin{equation}\label{21}
\epsilon_{3}=\frac{\mathcal{A}}{\mathcal{B}},
 \end{equation}
 and
 \begin{equation}\label{22}
\epsilon_{4}=\mathcal{C}+\mathcal{D},
 \end{equation}
 where
 \begin{eqnarray}\label{21-1}
\mathcal{A}&=&\big(2\beta H\dot{H}+24H\dot{H}\big)\alpha+\bigg[5+12^{n-2}\theta(n-1)n\big((2+\beta)H\dot{H}\big)^{n-2}+2\gamma\ln\big(12(2+\beta)H\dot{H}\big)\gamma\bigg],\nonumber\\
\mathcal{B}&=&2H\big(1+18\alpha(2+\beta)\big)H^{2}+\theta6^{n-1}\big((2+\beta)H^{2}\big)^{n-1}n+12(2+\beta)H^{2}\gamma\ln\big(6(2+\beta)H^{2}\gamma\big),\nonumber\\
\mathcal{C}&=&\frac{2\beta H^{2}\dot{H}+\dot{H}^{2}}{H \dot{H}},\nonumber\\
\mathcal{D}&=&\frac{12(2+\beta)\bigg[\frac{\alpha}{3(2+\beta)H^{2}}+\theta6^{n-3}
\big((2+\beta)H^{2}\big)^{n-3}(n-2)(n-1)n\bigg]\dot{H}}{1+18\alpha(2+\beta)H^{2}+\theta6^{n-1}\big((2+\beta)H^{2}\big)^{n-1}n+12(2+\beta)H^{2}\gamma\ln(6(2+\beta)H^{2}\gamma)}.
 \end{eqnarray}

Also, using the values obtained for slow-roll indices, $\epsilon_{i}, i=1...4$ and according to the Hubble rate given in  Eq. (\ref{18}) and with straightforward calculations as well as simplifications, the values of these indices can be obtained. Also, by using the values of slow-roll indices and even the Hubble rate, we get the scalar spectrum index (\ref{12}) and the tensor to the scalar ratio (\ref{14}). By performing some manipulations and then calculations, the variation rate of these variables concerning $n$ and $\beta$  are shown in the figures. We also describe these results. As you can see in figure \ref{1}, we plot the rate of change of the scalar-spectrum-index ($n_{s}$) in terms of the $\beta$ concerning the different values for the component of ($n$) and with respect to the constant values of the parameters such as $(\alpha)$, $(\theta)$ and $(\gamma)$. As you can see, the allowable range for this parameter is displayed and consistent with Planck's observable data. Also, the changes in this index in terms of (n) is well defined according to the parameter $(\beta)$ in Fig. \ref{2}. As shown in Fig. \ref{2} (b), for $(\beta=-1)$, the correct range of this index is displayed, which can be compared with the observable data.

\begin{figure}[h!]
 \begin{center}
 \subfigure[]{
 \includegraphics[height=4cm,width=4cm]{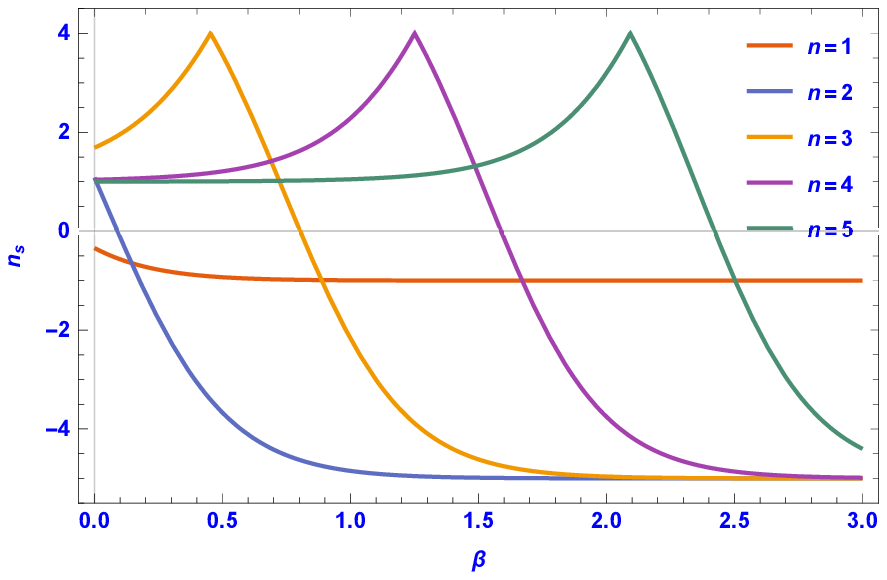}
 \label{1a}}
 \subfigure[]{
 \includegraphics[height=4cm,width=4cm]{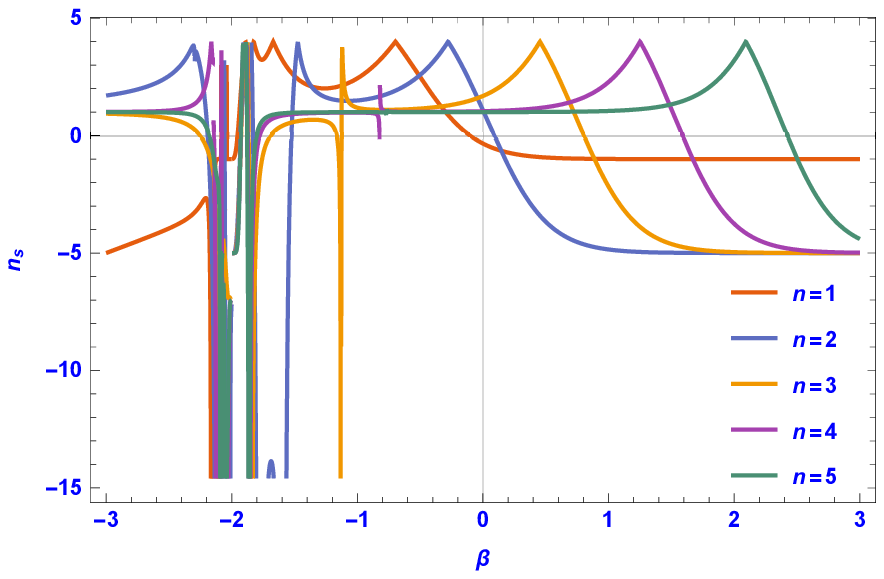}
 \label{1b}}
 \subfigure[]{
 \includegraphics[height=4cm,width=4cm]{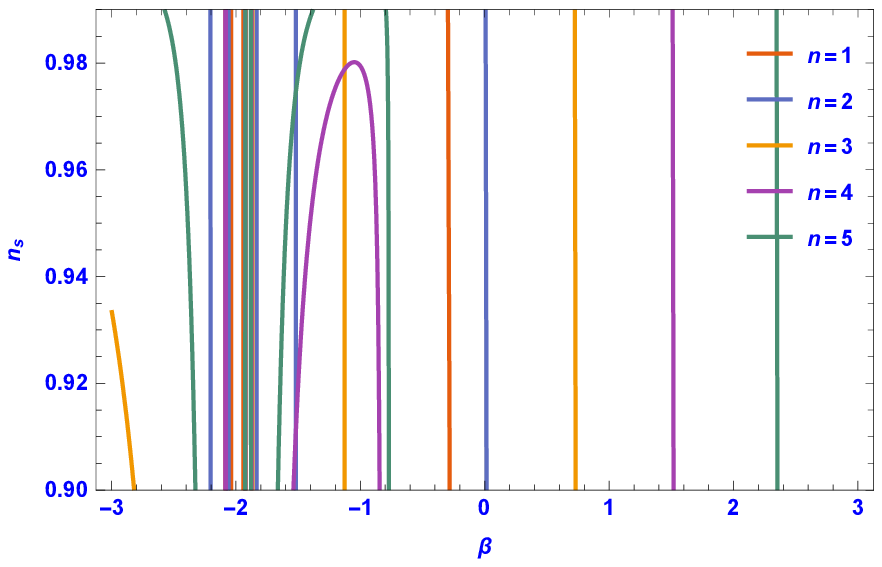}
 \label{1c}}
  \caption{\small{The plot of the variation of $n_{s}$ in terms of $0<\beta<3$ in the plot (a), $-3<\beta<3$ in the plots (b) and (c) with respect to different values of $n$ and the constant parameter $\alpha=0.15$, $\theta=0.009$ and $\gamma=0.01$. }}
 \label{1}
 \end{center}
 \end{figure}

\begin{figure}[h!]
 \begin{center}
 \subfigure[]{
 \includegraphics[height=5cm,width=5cm]{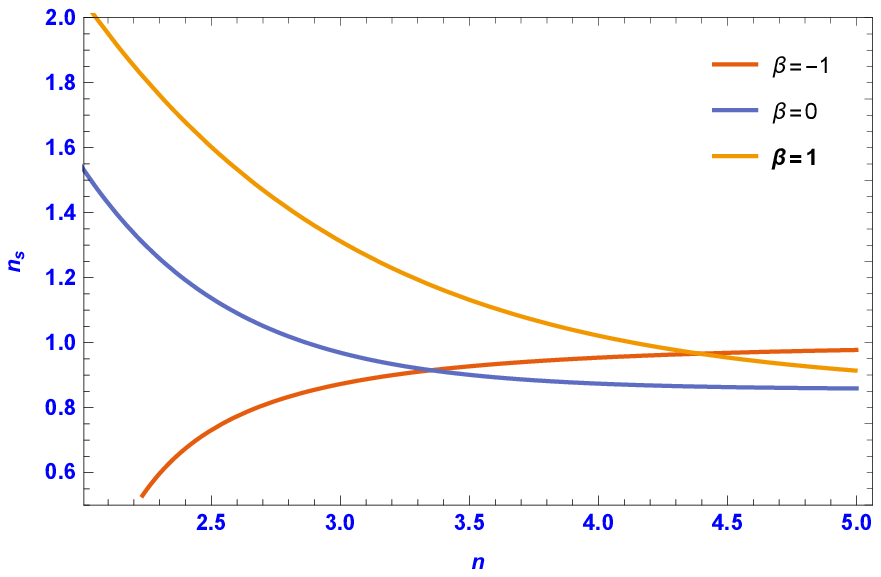}
 \label{2a}}
 \subfigure[]{
 \includegraphics[height=5cm,width=5cm]{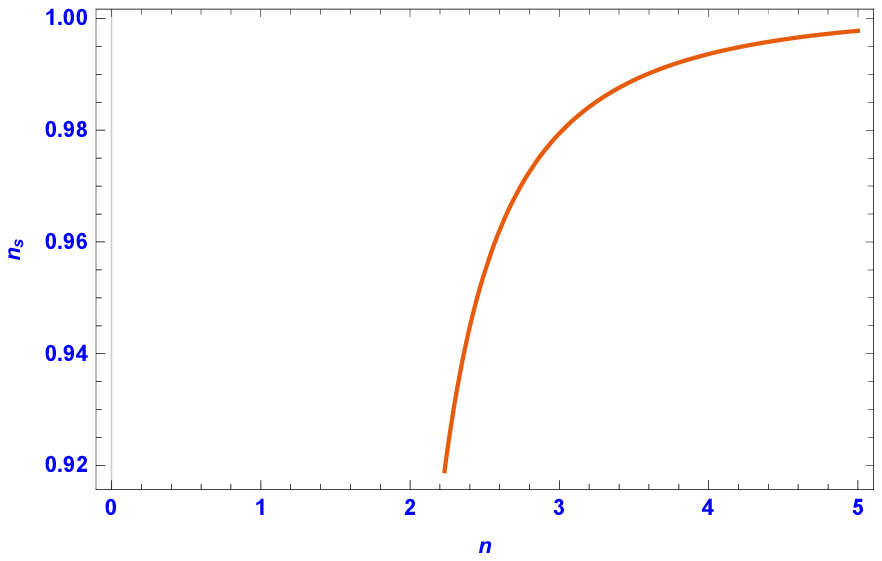}
 \label{2b}}
  \caption{\small{The plot of the variation of $n_{s}$ in terms of n and $-1<\beta<1$ in the plot (a) and $\beta=-1$ in the plot (b) with respect to constant parameter $\alpha=0.15$, $\theta=0.009$ and $\gamma=0.01$. }}
 \label{2}
 \end{center}
 \end{figure}

To understand the physical phenomena in the corresponding model, we take advantage of equations (\ref{12}), (\ref{17}), and (\ref{19})-(\ref{22}), and we plotted the different values of scalar spectrum index $n_{s}$ with respect to various parameters such as $\beta$ and $n$ Here,  we note that the variation rate in the above figures is comparable to the experimental data, especially Planck 2018 \cite{2}. Also, with respect to the above statement, As you can see in figure \ref{3}, We also plot the rate of change of the tensor-to-scalar ratio ($r$) in terms of the $\beta$ concerning the different values for the component of ($n$) and with respect to the constant values of the parameters such as $(\alpha)$, $(\theta)$ and $(\gamma)$. As you can see, the allowable range for this parameter is displayed in Fig. \ref{3} (a) for different values of $n$ and consistent with Planck's observable data.  Of course, in each case, the importance of these constant parameters are plotted by keeping the other one as a fixed parameter. Also as you can see in Fig. \ref{4}, we plot the rate of change of these two cosmological parameters, i.e., the scalar-spectrum-index ($n_{s}$) and the tensor-to-scalar ratio ($r$) to each other for different values of the parameters ($\beta$) and ($n$) for constant values ($\alpha$), ($\theta$) and ($\gamma$) as well as in the Figs. \ref{4} (a) and (b) are well determined the allowable range of these two cosmological parameters ($n_{s}$) and ($r$) proportional to each of the different values ($n$) and ($\beta$) are well specified.

 \begin{figure}[h!]
 \begin{center}
 \subfigure[]{
 \includegraphics[height=5cm,width=5cm]{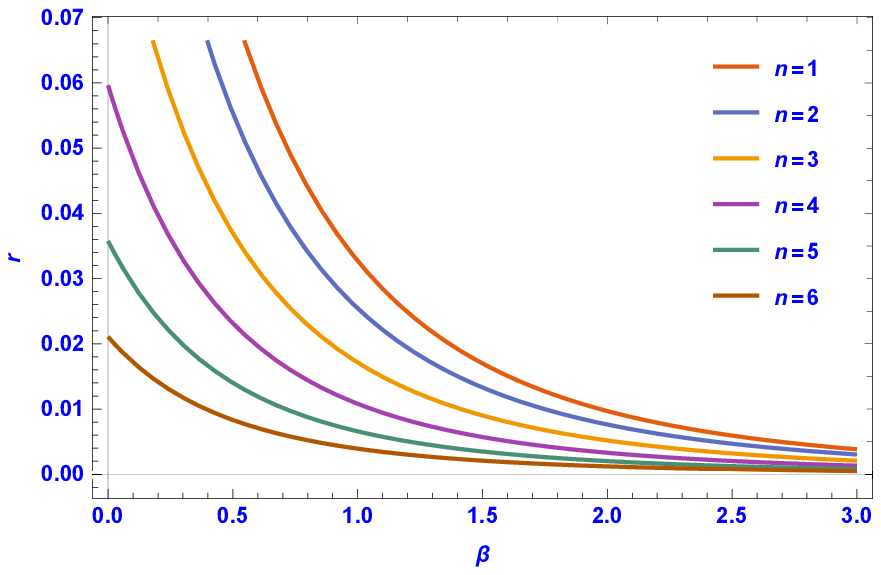}
 \label{3a}}
 \subfigure[]{
 \includegraphics[height=5cm,width=5cm]{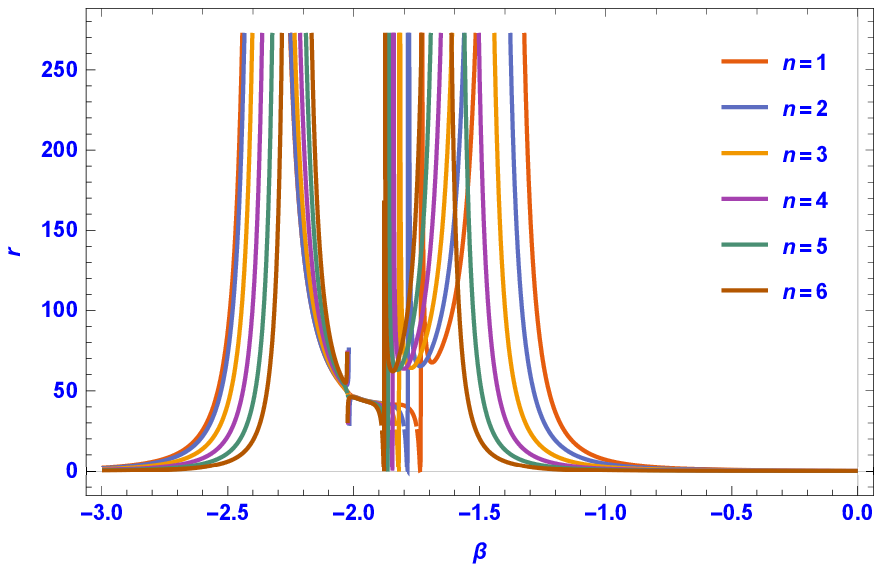}
 \label{3b}}
  \caption{\small{The plot of $r$ in terms of $0<\beta<3$ in the plot (a) and $-3<\beta<3$ in the plot (b) with respect to various values of $n$ and constant parameter $\alpha=0.15$, $\theta=0.009$ and $\gamma=0.01$. }}
 \label{3}
 \end{center}
 \end{figure}

\begin{figure}[h!]
 \begin{center}
 \subfigure[]{
 \includegraphics[height=4cm,width=4cm]{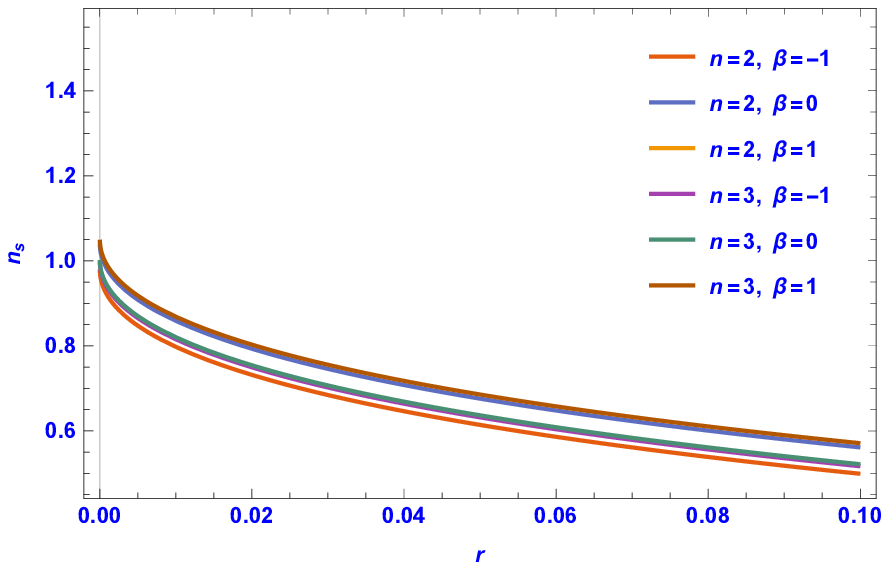}
 \label{4a}}
 \subfigure[]{
 \includegraphics[height=4cm,width=4cm]{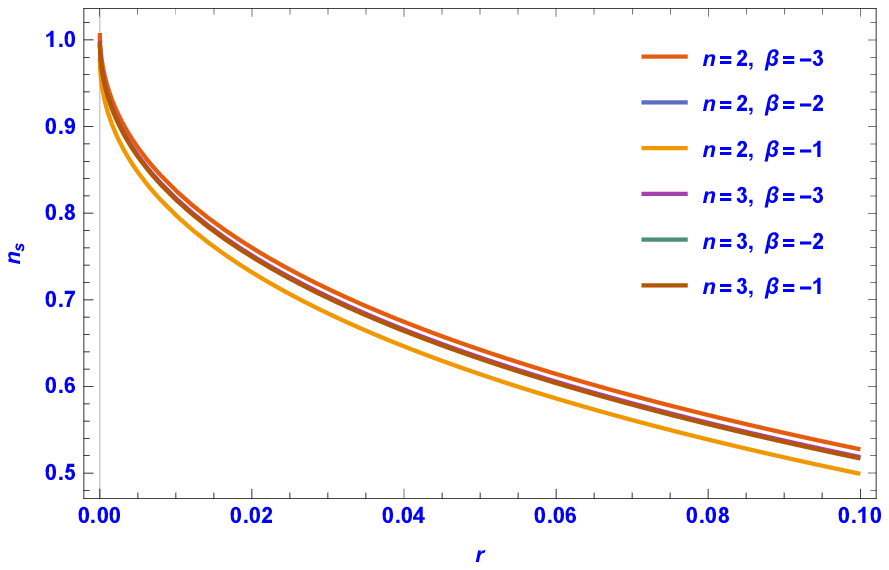}
 \label{4b}}
 \subfigure[]{
 \includegraphics[height=4cm,width=4cm]{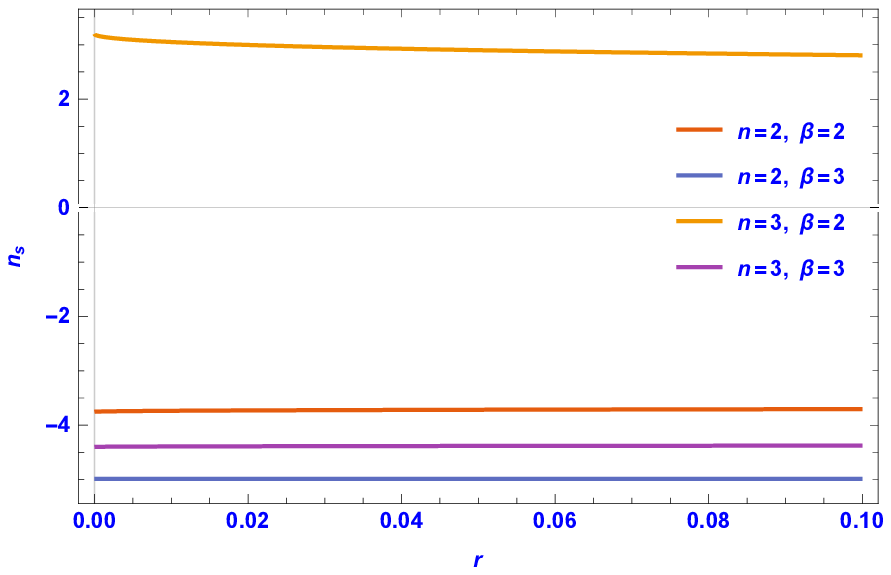}
 \label{4c}}
  \caption{\small{The $(n_{s}-r)$ plan with respect to different values of n and $\beta$ and the constant parameter $\alpha=0.15$, $\theta=0.009$ and $\gamma=0.01$.}}
 \label{4}
 \end{center}
 \end{figure}

Of course, as mentioned above, the modified $f(R)$ gravitational model has been studied under different constraints and conditions, such as the constant-roll and slow-roll conditions. And in this article, we take the general form $f(R)$ gravity. It has two exciting terms as polynomial and logarithmic form. It means that we examined the evolution of this model with several parameters. The logarithmic model is one of the essential inflation models that has been worked on by cosmologists in the last few years. As mentioned above,  the scalar spectrum index and the tensor-to-scalar ratio are only based on two constant parameters $\beta$ and $n$. A more detailed analysis shows that their values can be consistent with the observations for a large number of these parameters. We have given some examples in the form of the above figures.

\section{RSC in logarithmic constant roll inflation}
We seek to investigate the potential of this inflation model with respect to the assumption of weak gravity conjecture and the swampland condition from the point of view of constant-roll. As we know, the swampland condition is as follows \cite{20,21,22,23,24,25,26}.

\begin{equation}\label{23}
|\nabla V|\geq\frac{C_{1}}{M_{p}}V, \hspace{12pt} min(\nabla_{i}\nabla_{j}V)\leq -\frac{C_{2}}{M_{pl}^{2}}V
\end{equation}

The above equations for the $V>0$ can be rewritten in terms of the slow-roll parameters as follows,
\begin{equation}\label{04}
\sqrt{2\epsilon_{V}}\geq C_{1} ,\hspace{12pt}  or \hspace{12pt}\eta_{V}\leq -C_{2}
\end{equation}
where $C_1$ and $C_2$ are both positive and order of one, i.e., $C_1=C_2=\mathcal{O}(1)$. $ f(R)$ gravity is used for dark energy studies and cosmological models. As you know f is a function of the Ricci scale which $f = F + R$ \cite{51,52,53,54,55} used in cosmological studies and dark energy. Therefore, in this paper, we investigate this inflation model due to refined swampland conjecture and from the point of view of the constant-roll condition. In this case, we apply the modified $f(R)$ gravity and investigate inflationary theory in light of the above. Recently, a number of researchers have worked with the simple form of $f(R)$, which you see in Refs. \cite{56,57,58,59,60,61,62,63,64,65,66}. As we know, in general, the constant-roll condition for all types of inflation models, such as scalar field coupled to gravity and $f(R)$ gravitational models, has been investigated. Our goal is to investigate the $f(R)$ gravitational model using a constant-roll condition. There will be a special type of inflation solution for common equations of motion that we will briefly describe this route. Equations of motion which are given by,

\begin{equation}\label{24}
3H^{2}=\frac{\dot{\phi}}{2}+V(\phi),
 \end{equation}

\begin{equation}\label{25}
\frac{dH}{d\phi}=-\frac{\dot{\phi}}{2},
 \end{equation}
and
\begin{equation}\label{26}
 \ddot{\phi}+3H\dot{\phi}+\frac{\partial V}{\partial \phi }.
\end{equation}

The second derivative $\ddot{\phi}$ is negligible compared to the other equations above, which is ignored. More precise approximations, and some more precise solutions that give more accurate answers, have been used extensively in recent examples, especially when we faced with specific inflationary potentials $V(\phi)$ that have several non-analytical features \cite{67,68}. If $\frac{\partial V}{\partial \phi }$ is retained for a long time, which the second example refers to an ultra-slow-roll model. In agreement with the above equations, Einstein Hilbert action is investigated using a canonical scalar field as used in previous works \cite{69,70,71}. Similarly, in $f(R)$, gravitational models are a natural generalization of the constant-roll condition.

\begin{equation}\label{27}
\ddot{F}=\beta H \dot{F},
\end{equation}

where $\beta =-(3+\alpha)$. $\alpha$ is a non-zero parameter, and for $\alpha = -3$, the model is reduced to the standard slow-roll. Moving beyond the slow-roll approximation, we can consider an ultra slow-roll regime where the $\ddot{\phi}$ is finite in the Klein-Gordon equation as $\ddot{\phi}=3H\dot{\phi}$. As we know, this condition is in complete agreement with the previous one used in GR, and it is no different from the previous one. However, its generalization was stated in the preceding equation is very significant in many respects, and we only use the constant-roll condition for $f(R)$ gravitational models. using the constant-roll condition with respect to the refined swampland conjecture, and by applying them, we investigate the coefficient of swampland conjecture with logarithmic inflation model, we obtain some parameters such as the potential by Using the above equations, as well as experimental data and Planck 2018 data. Then we analyze the result of inflation model. Different types of inflation models from several perspectives such as slow roll, ultra-slow-roll, constant-roll (using methods such as beta and first-order function), as well as swampland program, including swampland conjectures, etc., have been studied, which examples of them were mentioned in the above remarks. first We calculate the potential, and then upper bound n. We are comparing our inflation model by plotting some figures. We also check to see if this gravitational model is consistent with swampland conjectures?  So first, we calculate the the potential; by using the Hubble parameter equations (\ref{18}), (\ref{1}), and (\ref{2}) with respect  to the $f (R)$ constant roll condition. Therefore one can obtain the final relation for the potential, which is given by,

 \begin{equation}\label{28}
 \begin{split}
&V(\phi)=\bigg\{-(\gamma^{2}\exp\big(\frac{n(24+12\alpha\beta+12^{n}(2+\beta)^{n}\gamma n)(144(2+\beta)^{2}(2^{1+2n}\times3^{n}\theta(2+\beta)^{n}+3n)+\phi)}{72(2+\alpha\beta)^{2}(2^{1+2n}\times3^{n}\alpha(2+\beta)^{n}+3n)}\big)\\
&\times n^{4}(24+12\alpha\beta+12^{n}\theta(2+\beta)^{n}n)^{2}(24+12\alpha\beta+12^{n}(2+\beta)^{n}\gamma n)^{2})+3\gamma^{2}n^{2}(24+12\alpha\beta+12^{n}\theta(2\beta)^{n}n)^{2}\\
&\times\bigg\{10368(2+\alpha\beta)^{4}(2^{1+2n}\times3^{n}\alpha(2+\beta)^{n}+3n)^{2}\bigg[+24n+2(6\alpha\beta n+6^{n}\theta(2+\beta)^{n}\gamma(-36\alpha\beta(2+\beta)^{2}n\\
&-(1+\beta)(n)^{2}))+\exp\big(\frac{n(24+12\alpha\beta+12^{n}(2+\beta)^{n}\gamma n)(144(2+\beta)^{2}(2^{1+2n}\times3^{n}\theta(2+\beta)^{n}+3n)+\phi)}{72(2+\alpha\beta)^{2}(2^{1+2n}\times3^{n}\alpha(2+\beta)^{n}+3n)}\big)\bigg]^{2}\bigg\}\bigg\}\\
&\bigg/\bigg(10368(2+\alpha\beta)^{4}(2^{1+2n}\times3^{n}\alpha(2+\beta)^{n}+3n)^{2}\bigg[+24n+2(6\alpha\beta n+6^{n}\theta(2+\beta)^{n}\gamma(-36\alpha\beta(2+\beta)^{2}n\\
&-(1+\beta)(n)^{2}))+\exp\big(\frac{n(24+12\alpha\beta+12^{n}(2+\beta)^{n}\gamma n)(144(2+\beta)^{2}(2^{1+2n}\times3^{n}\theta(2+\beta)^{n}+3n)+\phi)}{72(2+\alpha\beta)^{2}(2^{1+2n}\times3^{n}\alpha(2+\beta)^{n}+3n)}\big)\bigg]^{4}\bigg),
\end{split}
\end{equation}

After calculating the potential by using the constant-roll condition, we want consider two conditions of swampland conjectures, according to equation (\ref{23}), to determine whether the potential obtained from the above condition is in agreement with them. So with respect to  equation (\ref{28}), we need the first and second derivative of potential. We challenge the potential changes as well as the swampland conjecture by plotting some figures, and we will discuss the compatibility or incompatibility of the mentioned model with the swampland conjecture. As shown in figure \ref{5}, from left to right, the potential changes, the first and the second component of the swampland conjecture as $C_{1}$ and $C_{2}$  are plotted according to the scalar field $\phi$ and different values of the constant parameter $\alpha$, $\theta$ and $\gamma$.
The changes of each of these quantities regarding the constant roll condition parameter $\beta$ are shown. In the literature, components $C_{1}$ and $C_{2}$ are usually constant, positive, and the unit order that the second component $C_{2}$ has smaller values than $C_{1}$.
 As it is clear from the figure \ref{5}, the first and second components of the swampland conjectures are in their desired range, and also the change of these two component for the various values of scalar field $\phi$ and the constant parameter $\beta$ is well known.
Also,  shown in Fig. \ref{5}, the $C_{2}$ has smaller value than the $C_{1}$, and a kind of optimal compatibility of these different conditions is seen.

\begin{figure}[h!]
 \begin{center}
 \subfigure[]{
 \includegraphics[height=4cm,width=4cm]{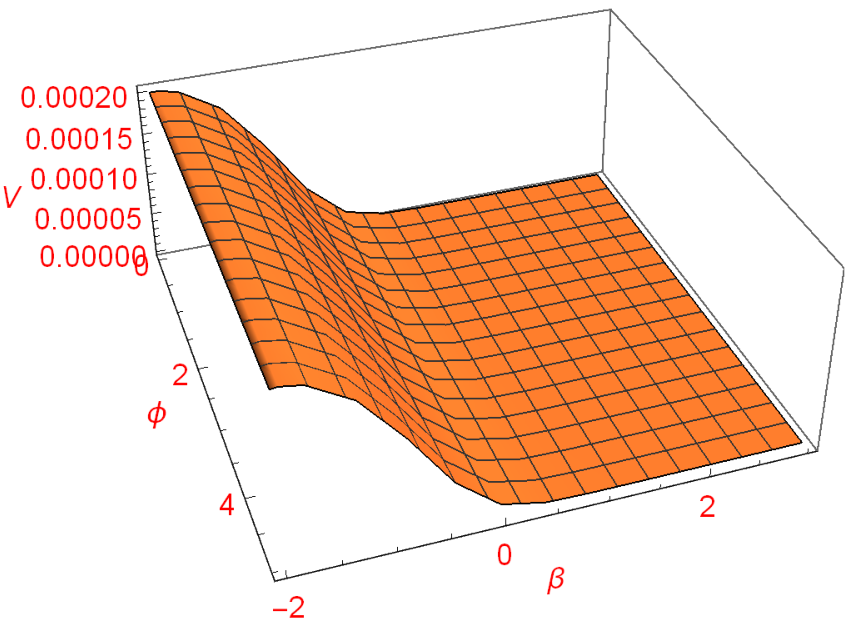}
 \label{5a}}
 \subfigure[]{
 \includegraphics[height=4cm,width=4cm]{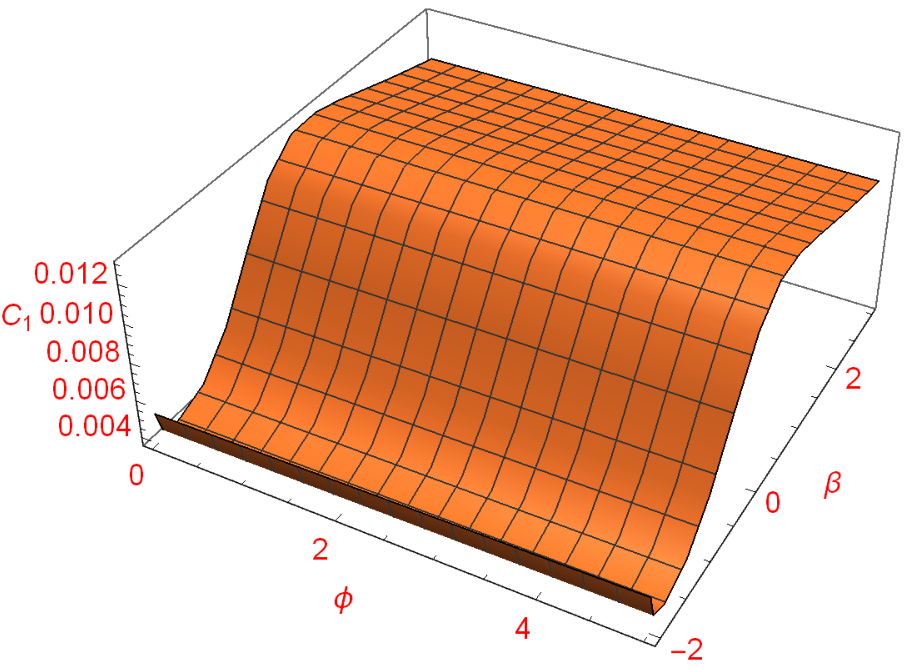}
 \label{5b}}
 \subfigure[]{
 \includegraphics[height=4cm,width=4cm]{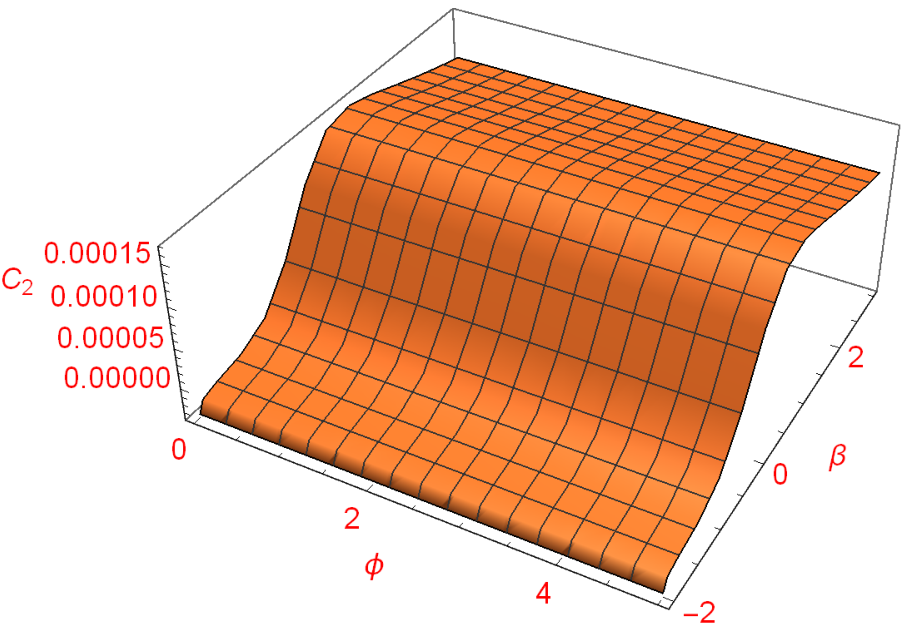}
 \label{5c}}
  \caption{\small{The plot of $V$, $C_{1}$, and $C_{2}$ in term of $\phi$ with respect to different values of $\beta$, and the constant parameter $\alpha=0.15$, $\theta=0.009$ and $\gamma=0.01$.}}
 \label{5}
 \end{center}
 \end{figure}

Nevertheless, in all the above calculations, it is useful these limits to calculations and figures. In this article, we have used one of the conditions for  swampland program. Today, other conditions related to the swampland whies are more powerful conjectures such as trans-Plankcian conjecture (TCC) used to investigate the inflation models with certain restrictions. Each of the inflationary models can be examined and studied from the point of view of these conditions.

\section{Conclusions}
A more general and special form of an inflationary model called the constant-roll condition has been replaced by two-parameter phenomenological inflationary models in $GR$ in the slow-roll condition, which was used to study generalized gravitational $f(R)$ models. In this paper, we studied the logarithmic $f(R)$ cosmic evolution with respect to refined swampland conjecture. To review the above, we first introduced our inflationary model, i.e., logarithmic $f(R)$ gravitational model, which is a polynomial function with a logarithmic term. Then we explained the constant-roll model. We investigated the logarithmic inflation model using constant-roll conditions (constant rate of inflation), and we obtained values such as potential and  Hubble parameter. We know that the potential value obtained with this condition has an exact value. Also We challenged our inflation model with respect to refined swampland conjectures. We concluded that these conditions challenged the swampland conjecture. We examined the constant-roll evolution with logarithmic $ f (R) $ gravity. With the constant-roll conditions (\ref{17}) and performing some manipulations with straightforward calculations and simplifications, we achieved the final relation for the Hubble parameter $H$. This relation helped us to have more information about the corresponding system.  After then we plotted figures such as $n_{s}$  with respect to $n$ and $\beta$ separately. Also, we plot the figures $r$ concerning $n$ and $\beta$ and model's constant parameters, i.e., $\alpha$, $\theta$ and $\gamma$ respectively. In that case, we had some suitable results, which is explained by several figures. Finally,  we analyzed the figures and evaluated the calculations obtained with respect to the experimental data, especially Planck 2018 \cite{2}.


\begin{thebibliography}{11}
\bibitem{1}
M. Jerome, R. Christophe, and V. Vincent, "Encyclopdia Inflationaris", Phys. Dark Univ 5-6, 75-235 (2014).
\bibitem{2}
Y. Akrami et al., "Planck 2018 results-X. Constraints on inflation",  A\&A 641, A10 (2020).
\bibitem{3}
S. Nojiri, S. D. Odintsov, "Unified cosmic history in modified gravity: from F (R) theory to Lorentz non-invariant models", Phys. Rept. 505, 59 (2011).
\bibitem{4}
S. Nojiri, S.D. Odintsov, "Introduction to modified gravity and gravitational alternative for dark energy", Int. J. Geom. Meth. Mod. Phys. 4, 115 (2007).
\bibitem{5}
S. Capozziello, M. De Laurentis, "Extended theories of gravity", Phys. Rept. 509, 167 (2011).
\bibitem{6}
R. Myrzakulov, L. Sebastiani and S. Zerbini, "Some aspects of generalized modified gravity models", Int. J. Mod. Phys. D 22, 1330017 (2013).
\bibitem{7}
S. Nojiri, S. D. Odintsov, and V. K. Oikonomou, "Modified gravity theories on a nutshell: inflation, bounce and late-time evolution", Phys. Rept. 692, 1-104 (2017).
\bibitem{8}
S. Nojiri and S. D. Odintsov, "Modified gravity with negative and positive powers of curvature: Unification of inflation and cosmic acceleration", Phys. Rev. D68, 123512 (2003).
\bibitem{9}
G. Cognola, E. Elizalde, S. Nojiri, S. D. Odintsov, L. Sebastiani and S. Zerbini, "Class of viable modified  gravities describing inflation and the onset of accelerated expansion", Phys. Rev. D 77, 046009 (2008).
\bibitem{10}
E. Elizalde, S. Nojiri, S. D. Odintsov, L. Sebastiani and S. Zerbini, "Nonsingular exponential gravity: a simple theory for early-and late-time accelerated expansion", Phys. Rev. D 83, 086006 (2011).
\bibitem{11}
J. Martin, H. Motohashi, and T. Suyama, "Ultra slow-roll inflation and the non-Gaussianity consistency relation", Phys. Rev. D87, 023514 (2013).
\bibitem{12}
H. Motohashi, A. A. Starobinsky, and J. Yokoyama, "Inflation with a constant rate of roll", JCAP 1509, 018 (2015).
\bibitem{13}
H. Motohashi, and A. A. Starobinsky, "Constant-roll inflation: confrontation with recent observational data", Europhys. Lett. 117, 39001 (2017).
\bibitem{14}
N. Rashidi, M. Heidarzadeh, and K. Nozari, "Constant-roll inflation with hilltop potential", The European Physical Journal Plus137, 514 (2022).
\bibitem{15}
M. Shokri, J. Sadeghi, and S. Noori Gashti, "Quintessential constant-roll inflation", Physics of the Dark Universe
35, 100923 (2022).
\bibitem{16}
M. H. Namjoo, H. Firouzjahi, and M. Sasaki, "Violation of non-Gaussianity consistency relation in a single-field inflationary model", Europhys.Lett. 101, 39001 (2013).
\bibitem{166}
I. Antoniadis, A. Lykkas, and K. Tamvakis, "Constant-roll in the Palatini-R2 models", JCAP 04, 033 (2020).
\bibitem{17}
M. Orellana, F. Garcia, F. Teppa Pannia, and G. Romero, "Structure of neutron stars in R-squared gravity", Gen. Rel. Grav 45, 771-783 (2013).
\bibitem{18}
S. Capozziello, M. De Laurentis, S. D. Odintsov, and A. Stabile, "Hydrostatic equilibrium and stellar structure in  gravity", Phys. Rev. D 83, 064004 (2011).
\bibitem{19}
S. Capozziello, M. Faizal, M. Hameeda, B. Pourhassan, V. Salzano, and S. Upadhyay, "Clustering of galaxies with f(R) gravity", Mon. Not. Roy. Astron. Soc. 474, 2430-2443 (2018).
\bibitem{20}
A. Arapoglu, C. Deliduman, and K. Y. Eksi, "Constraints on perturbative f (R) gravity via neutron stars", JCAP 1107, 020 (2011).
\bibitem{21}
S. Capozziello, R. D'Agostino, and O. Luongo, "Extended gravity cosmography", Int. J. Mod. Phys. D, 28, 1930016 (2019).
\bibitem{22}
S. Capozziello, R. D'Agostino, and O. Luongo, "Rational approximations of f (R) cosmography through Pad'e polynomials", JCAP 1805, 008 (2018).
\bibitem{23}
M. Khurshudyan, A. Pasqua, and B. Pourhassan, "Higher derivative corrections of f(R) gravity with varying equation of state in the case of variable G and $\Lambda$", Can. J. Phys. 1107, 449-455 (2015).
\bibitem{24}
Ph. Channuie, "Deformed Starobinsky model in gravity's rainbow", Eur. Phys. J. C, 79, 508 (2019).
\bibitem{25}
S. Capozziello, R. D'Agostino, and O. Luongo, "Kinematic model-independent reconstruction of Palatini f (R) cosmology", Gen. Rel. Grav. 51, 2 (2019).
\bibitem{27}
R. Myrzakulov, L. Sebastian, and S. Vagnozzi, "Inflation in f( R , $\phi$ )-theories and mimetic gravity scenario", Eur. Phys. J.C 75, 444 (2015).
\bibitem{28}
J. Sadeghi, B. Pourhassan, A. S. Kubeka, and M. Rostami, "Logarithmic corrected polynomial f(R) inflation mimicking a cosmological constant", Int. J. Mod. PhysD, 25, 1650077 (2015).
\bibitem{29}
S. Nojiri and S. D. Odintsov, "Modified gravity with ln R terms and cosmic acceleration", Gen. Rel. Grav. 36, 1765 (2004).
\bibitem{30}
S. D. Odintsov, V. K. Oikonomou, and L. Sebastiani, "Unification of constant-roll inflation and dark energy with logarithmic R2-corrected and exponential $F(R)$ gravity", Nucl. Phys. B, 923, 608 (2017).
\bibitem{31}
E. Elizalde, S. D. Odintsov, L. Sebastiani, and R. Myrzakulov, " Beyond-one-loop quantum gravity action yielding both inflation and late-time acceleration", Nucl. Phys. B, 921, 411 (2017).
\bibitem{32}
J. L. Cook, and L. M. Krauss, "Large slow roll parameters in single field inflation", JCAP 03, 028 1603 (2016).
\bibitem{33}
K. S. Kumar, J. Marto, P. Vargas Moniz, and S. Das, "Non-slow-roll dynamics in $\alpha$-attractors", JCAP 04, 005 1604 (2016).
\bibitem{34}
S. D. Odintsov, and V. K. Oikonomou, "Inflationary dynamics with a smooth slow-roll to constant-roll era transition", JCAP 04, 041 1704 (2017).
\bibitem{35}
S. D. Odintsov, and V. K. Oikonomou, "Inflation with a smooth constant-roll to constant-roll era transition", Phys. Rev. D 96 2, 024029 (2017).
\bibitem{36}
Q. Gao et al., "Constant-roll tachyon inflation and observational constraints", JCAP 05, 005 (2018).
\bibitem{37}
Q. Gao, and Y. Gong, "Reconstruction of extended inflationary potentials for attractors", Eur. Phys. J. Plus 133 11, 491(2018).
\bibitem{38}
Q. Gao, "Reconstruction of constant slow-roll inflation", Sci. China Phys. Mech. Astron. 60 9, 090411 (2017).
\bibitem{39}
Q. Fei, Y. Gong, J. Lin, and Z. Yi, "The reconstruction of tachyon inflationary potentials", JCAP 08  018 (2017).
\bibitem{jhap}
S. Noori Gashti, "Two-field inflationary model and swampland de Sitter conjecture", Journal of Holography Applications in Physics 2(1) 13-24 (2022). doi: 10.22128/jhap.2021.452.1002
\bibitem{jhap2}
S. Maity, P. Rudra, "Inflation driven by Barrow holographic dark energy", Journal of Holography Applications in Physics 2(1) 1-12 (2022). doi: 10.22128/jhap.2022.464.1012
\bibitem{26}
J. Sadeghi, E. Naghd Mezerji, and S. Noori Gashti, "Study of some cosmological parameters in logarithmic corrected  gravitational model with swampland conjectures", Mod. Phys. Lett. A (2020).
\bibitem{266}
J. Sadeghi, and S. Noori Gashti, "Anisotropic constant-roll inflation with noncommutative model and swampland conjectures", Eur. Phys. J. C 81, 301 (2021).
\bibitem{2666}
J. Sadeghi, S. Noori Gashti, and E. Naghd Mezerji, "The investigation of universal relation between corrections to entropy and extremality bounds with verification WGC", Phys. Dark Univ 30, 100626, (2020).
\bibitem{311}
S. K. Garg, and C. Krishnan, "Bounds on slow roll and the de Sitter swampland", J. High Energ. Phys. 2019, 75 (2019).
\bibitem{3111}
V. K. Oikonomou, "Rescaled Einstein-Hilbert gravity from  gravity: Inflation, dark energy, and the swampland criteria", Phys. Rev. D 103, 124028 (2021).
\bibitem{110}
J. Sadeghi, S. N. Gashti, and F. Darabi, "Swampland conjectures in hybrid metric-Palatini gravity", Physics of the Dark Universe, 37, 101090 (2022).
\bibitem{111}
A. Castellano, A. Herr\'{a}ez, and L. E. Ib\'{a}\~{n}ez, "IR/UV mixing, towers of species and swampland conjectures", Journal of High Energy Physics 2022, 217 (2022).
\bibitem{112}
S. Noori Gashti, and J. Sadeghi, "Refined swampland conjecture in warm vector hybrid inflationary scenario",
The European Physical Journal Plus137, 731 (2022).
\bibitem{113}
J. Yuennan, and P. Channuie, "Composite Inflation and further refining dS swampland conjecture", arXiv:2208.09842 (2022).
\bibitem{114}
R. Alvarez-García, R. Blumenhagen, and C. Kneil, "Swampland conjectures for an almost topological gravity theory", Physics Letters B 825, 10 136861 (2022).
\bibitem{115}
F. B. M dos Santos, R. Silva, and S. S. da Costa, "Warm $\beta$ -exponential inflation and the Swampland Conjectures",  arXiv:2209.06153v1 (2022).
\bibitem{116}
S. N. Gashti, J. Sadeghi, and B. Pourhassan, "Pleasant behavior of swampland conjectures in the face of specific inflationary models", Astroparticle Physics 139, 102703 (2022).
\bibitem{117}
S. N. Gashti, J. Sadeghi, "Constraints on cosmological parameters in light of the scalar–tensor theory of gravity and swampland conjectures", Modern Physics Letters A 37 (18), 2250110 (2022).
\bibitem{118}
E. Gonzalo, L. E. Ibáñez, and I. Valenzuela, "Swampland constraints on neutrino masses", Journal of High Energy Physics2022, 88 (2022).
\bibitem{119}
O. Bertolami, P. M. Sa, "Multi-field cold and warm inflation and the de Sitter swampland conjectures", JCAP 09,  001 (2022).
\bibitem{120}
R. K. Mishra, "Confinement in de Sitter Space and the Swampland", arXiv:2207.12364, (2022).
\bibitem{121}
A. Mohammadi, T. Golanbari, S. Nasri, and Kh. Saaidi, "Brane inflation: Swampland criteria, TCC, and reheating predictions", Astroparticle Physics142, 102734 (2022).
\bibitem{122}
D. Andriot, L. Horer, "(Quasi-) de Sitter solutions across dimensions and the TCC bound", arXiv:2208.14462, (2022).
\bibitem{123}
J. Sadeghi, M. R. Alipour, S. N. Gashti, "Scalar Weak Gravity Conjecture and Inflationary Models", arXiv:2208.13093, (2022).
\bibitem{40}
S. Nojiri, S. D. Odintsov, and V. K. Oikonomou, "Constant-roll inflation in F (R) gravity", Class.Quant.Grav. 34 24, 245012 (2017).
\bibitem{41}
H. Motohashi, and A. A. Starobinsky, "f (R) constant-roll inflation", Eur. Phys. J. C 77 8, 538 (2017).
\bibitem{b}
S. Nojiri, and S. D. Odintsov, "Modified gravity with ln R terms and cosmic acceleration", Gen. Rel. Grav. 36, 1765 (2004).
\bibitem{d}
Qing-Guo Huang, "A polynomial $f(R)$ inflation model", JCAP 02, 035 (2014).
\bibitem{e}
T. Saidov, A. Zhuk, "Bouncing inflation in nonlinear $R^{2} + R^{4}$ gravitational model", Phys. Rev. D81 124002 (2010).
\bibitem{f}
J. Sadeghi, H. Farahani, "Logarithmic corrected F (R) gravity in the light of Planck 2015", Physics Letters B 751, 89-95 172015 (2015).
\bibitem{g}
J. Sadeghi, and S. Noori Gashti, "Investigating the logarithmic form of f (R) gravity model from brane perspective and swampland criteria", Pramana 95 (198) (2021).
\bibitem{h}
J. Sadeghi, M. Shokri, S. Noori Gashti, B. Pourhassan, and P. Rudra, "Traversable wormhole in logarithmic  gravity by various shape and redshift functions", Int. J. Mod. Phys. D, 2250019 (2022).
\bibitem{42}
V. K. Oikonomou, "Reheating in Constant-roll F(R) Gravity", Mod. Phys. Lett. A  33, 1750172 (2017).
\bibitem{43}
H. Noh, and J. c. Hwang, "Inflationary spectra in generalized gravity: Unified forms", Phys. Lett. B 515 (2001).
\bibitem{44}
J. c. Hwang, and H. r. Noh, "Gauge-ready formulation of the cosmological kinetic theory in generalized gravity theories", Phys. Rev. D 65, 023512 (2002).
\bibitem{45}
J. c. Hwang and H. Noh, "f (R) gravity theory and CMBR constraints", Phys. Lett. B 506 (2001).
\bibitem{46}
S. Nojiri, S. D. Odintsov, and V. K. Oikonomou, "Viable mimetic completion of unified inflation-dark energy evolution in modified gravity", Phys. Rev. D 94 10, 104050 (2016).
\bibitem{47}
S. D. Odintsov, and V. K. Oikonomou, "Singular F (R) cosmology unifying early-and late-time acceleration with matter and radiation domination era", Class. Quant. Grav. 33 12, 125029 (2016).
\bibitem{48}
S. D. Odintsov, and V. K. Oikonomou, "Singular inflationary universe from  gravity", Phys. Rev. D 92 12, 124024 (2015).
\bibitem{50}
M. Guerrero, D. Rubiera-Garcia, and D. Saez-Chillón Gomez, "Constant roll inflation in multifield models", Phys. Rev. D 102, 123528 (2020).
\bibitem{51}
A. A. Starobinsky, "A new type of isotropic cosmological models without singularity", Phys. Lett. B, 91, 99-102 (1980).
\bibitem{52}
A. Codello, J. Joergensen, F. Sannino, and O. Svendsen, "Marginally deformed Starobinsky gravity", JHEP 02, 050 (2015).
\bibitem{53}
S. Capozziello, M. De Laurentis, R. Farinelli, and Sergei D. Odintsov, "Mass-radius relation for neutron stars in
f(R) gravity", Phys. Rev. D 93, 023501 (2016).
\bibitem{54}
K. Bamba et al., "Bounce cosmology from F (R) gravity and F (R) bigravity", JCAP 01, 008 (2014).
\bibitem{55}
G. J. Olmo, D. Rubiera-Garcia, and A. Wojnar, "Stellar structure models in modified theories of gravity: Lessons and challenges", Physics Reports 876, 13  1-75 (2020).
\bibitem{56}
S. Capozziello, E. De Filippis, and V. Salzano, "Modelling clusters of galaxies by f(R) gravity", Monthly Notices of the Royal Astronomical Society, 394, 2, 947–959 (2009).
\bibitem{57}
A. Arapoglu, C. Deliduman and K. Y. Eksi, "Constraints on perturbative f (R) gravity via neutron stars", JCAP 1107, 020 (2011).
\bibitem{58}
S. Capozziello, and M. De Laurentis, "Extended Theories of Gravity", Physics Reports509, 4–5, 167-321 (2011).
\bibitem{59}
Chao-Qiang Geng et al., "Cosmological Constraints on Nonflat Exponential f(R) Gravity", ApJ 926, 74 (2022).
\bibitem{60}
S. D. Odintsov, and V. K. Oikonomou, "Pre-inflationary bounce effects on primordial gravitational waves of f(R) gravity", Physics Letters B824, 10 136817 (2022).
\bibitem{61}
Ph. Channuie, Eur. Phys. J. C, 79, 508 (2019).
\bibitem{62}
Y. Leyva, and G. Otalora, "Revisiting  gravity's rainbow: Inflation and primordial fluctuations", arXiv:2206.09000, (2022).
\bibitem{63}
V.K. Oikonomou, "Kinetic axion $F(R)$ gravity inflation", Phys. Rev. D 106, 044041 (2022).
\bibitem{64}
S. Noori Gashti et al., "Swampland dS conjecture in mimetic $f(R, T)$ gravity", Commun. Theor. Phys. 74 085402 (2022 ).
\bibitem{65}
M. Koussour, S. H. Shekh, A. Hanin, Z. Sakhi, S. R. Bhoyer, and M. Bennai, "Flat FLRW Universe in logarithmic symmetric teleparallel gravity with observational constraints",  arXiv:2203.00413v2 (2022).
\bibitem{66}
L. V. Jaybhaye, S. Mandal, and P. K. Sahoo, "Constraints on energy conditions in $f(R,Lm)$ gravity", International Journal of Geometric Methods in Modern PhysicsVol. 19, No. 04, 2250050 (2022).
\bibitem{67}
A. K. Sharma, and M. M. Verma, "Power-law Inflation in the $f(R)$ Gravity", ApJ 926, 29 (2022).
\bibitem{68}
S. Inoue and J. Yokoyama, "urvature perturbation at the local extremum of the inflaton potential", Phys.Lett. B524, 15 (2002).
\bibitem{69}
J. Martin, H. Motohashi, and T. Suyama, "Ultra slow-roll inflation and the non-Gaussianity consistency relation", Phys. Rev. D87, 023514 (2013).
\bibitem{70}
T. J. Gao, "Gauss-Bonnet inflation with a constant rate of roll",  The European Physical Journal C 80, 1013 (2020).
\bibitem{71}
A. Karam et al., "Constant-roll (quasi-)linear inflation", JCAP05, 011 (2018).

\end{thebibliography}
\end{document}